\documentclass[aps,prl,twocolumn,showpacs,superscriptaddress,groupedaddress]{revtex4}
\usepackage{graphicx}  
\usepackage{dcolumn}   
\usepackage{bm}        
\usepackage{amssymb}   
\usepackage{slashed}
\usepackage{graphicx}				
\usepackage{amsmath}
\usepackage{mathtools}
\usepackage{tikz}
\usetikzlibrary{arrows,backgrounds}
\usepackage[all]{xy}
\usepackage{yfonts}
\begin{document}

\title{On quantization and general relativity}
\author{Andrei T. Patrascu}
\address{University College London, Department of Physics and Astronomy, London, WC1E 6BT, UK}
\begin{abstract}
This note is to bring to the reader's attention the fact that general relativity and quantum mechanics differ from each other in one main aspect. General relativity is based on the diffeomorphism covariant formulation of the laws of physics while quantum mechanics is constructed such that its fundamental laws remain invariant to a change of topology. It is the goal of this paper to show that in order to obtain a complete description of quantum gravity one has to extend the principle of diffeomorphism invariance from general relativity in the sense of quantum mechanics i.e. the laws of physics must be covariant to a change in the topology of spacetime. 
\end{abstract}
\maketitle
\section{Introduction}
The prescriptions of general relativity and quantum mechanics are taking away most of the absoluteness associated to choices of coordinates, trajectories followed by particles and states of physical systems in the absence of any accessible information about them. It is my observation that there still remains an epistemological defect associated to these ideas. Not to all arbitrary conventions has been taken their absolute status away. In fact the connectivity of space is probably the last convention that still is considered absolute by many physicists. It is my observation that one cannot assign a specific absolute topology to spacetime itself in the absence of a method of detecting such a topology. It is also my observation that in the absence of specific information related to the topology of space no such property can be rigorously associated to it without arriving at paradoxes and inconsistencies. Because of this, it appears to be necessary for the laws of nature to be specified in a topology-covariant way. This attempt is made in this article. 
\section{Covariance principles in physics}
The main developments of the past century in physics (special relativity, general relativity and quantum mechanics) have brought to our attention the fact that abstract mathematical conventions should not stand at the fundaments of physics. In general, the role of conventions is to facilitate the comprehension of physical reality and not to assign physical reality to conventional constructions. This idea was noted probably for the first time by Einstein and incorporated in his theory of special relativity as the weak equivalence principle: "the laws of nature should not depend on the arbitrary choice of an inertial reference frame". Of course, this law was further generalized to the statement that "the general laws of nature are to be expressed by equations which hold good for all systems of co-ordinates, that is, are co-variant with respect to any substitutions whatsoever (generally co-variant)"[1]. 
\par It was the goal of general relativity to eliminate any absolute character associated to spacetime coordinates and to refer to them as to arbitrary choices. Geometrical curvature of space-time was to be associated to what was before known as "force of gravity". Diffeomorphism invariance (i.e. invariance of the laws of physics under the smooth local shift of a path in space-time followed by a reparametrization to the initial numerical coordinate values) was formulated as a fundamental invariance of nature. Indeed, following this way of thinking there is no difference between an accelerated motion and a motion in a gravitational field as seen from a local standpoint. The result was the general theory of relativity which is certainly among the best tested constructions of theoretical physics. There is however a situation in which general relativity fails. The associated phenomenon is known as a "black hole" and is probably one of the most intensely studied phenomena today [2],[3]. A black hole is in general characterized, among other, by the area of its horizon. 
Physically, the horizon of a black hole is a classical surface of no return. It is assumed that it also gives non-trivial topology to the space-time that contains it. One can define a black hole as the region encapsulated by such a horizon. General relativity is predicted to fail at some point behind a horizon. However, generally, the connectivity of spacetime invariantly changes when a horizon forms. As probably I do not have to remind here, there is no possible causal connection between the interior of a black hole and the exterior. It results that a consequence of the formation of a black hole is the alteration of the topology (connectivity) of space-time. 
Let me now consider general diffeomorphism invariance in the presence of a black hole. Diffeomorphism invariance is assumed to be a symmetry of general relativity. That means, if one performs a push-forward of a path and then a coordinate transformation to the original numerical values of the coordinates one should obtain no change in the formal expression of the theory. However, one will encounter problems when dealing with topologically non-trivial spacetimes. One may observe that a hole in space will make some transformations impossible from purely geometrical reasons. In general relativity, there exist points that are associated to divergences of diffeomorphism invariant objects (scalars). These are the points where the prescriptions of general relativity fail.
\par One aspect has however been overlooked when dealing with this type of problems. In flat or curved spacetime without horizons the laws of physics have to be formulated in a covariant way (due to the strong equivalence principle) and it is essentially impossible to tell the difference between a curved space and an accelerating frame using only local measurements.
What about the choice of a topology? Should one be able in principle to infer the existence of a non-trivial topology in the absence of an experimental setup constructed such that the connectedness of spacetime becomes manifest? Otherwise stated is the topology of spacetime a fundamental absolute property of space? 
The universal coefficient theorems show that in principle there is nothing absolute related to connectedness and classes of homotopy or (co)homology and that one cannot expect to have a universal absolute topological setup over spacetime. 
Is there a physical reason for this lack of absoluteness? In fact, there are several paradoxes that appear due to the assumption that spacetime connectivity should have a special character. 
 
 The problem one is faced with is: should the laws of physics change dramatically in the presence of non-trivial topologies? Is there a special feature associated to a topologically non-trivial space that makes it special among all the other possible choices? On the other side, even in the absence of a macroscopic black hole, measurements of intervals in space-time are possible only if enough energy is concentrated in a small region. The more accurate measurements we want the more energy has to be added to the specific region. Finally one may add sufficient energy such that the spacetime will become topologically non-trivial. By this, the mere procedure of measuring distances may entail non-trivial topologies and consequently horizons and Hawking radiation that may be detected locally from far away. One may ask then, how is it possible to have singled out a whole category of spacetimes that depend only on our choices of performing measurements of some nature? 
Do the laws of nature change when working with a topologically non-trivial spacetime? 
\par Indeed it is the goal of this article to show that there are no special requirements for topologically non-trivial spacetimes and that the laws of nature are fundamentally independent of the connectivity assigned to a given space. In essence topology is just another word for connectedness. The scope of topology is to identify and measure connected structures in an abstract space. However, it was already clear for Einstein that "all our space-time verifications invariably amount to a determination of space-time coincidences. If for example, events consisted merely in the motion of material points, then ultimately nothing would be observable but the meetings of the material points of our measuring instruments with other material points (...)"[1]. 
It is at this point where the connection to quantum mechanics must be introduced. There, besides the immaterial character of an absolute spacetime reference, one adds the fact that specific intermediate states of particles (even intermediate positions) are not to be considered physical unless they are practically observed. As a result, following Feynman's path integral prescription [4], one is capable of formulating quantum mechanics in the form of functional integrals over field spaces. All possible paths that connect two events in space-time must be considered as possible alternatives in calculating the amplitudes. Born rules [5] will provide us with the probabilities for various events such that we will be able to construct the statistics. The calculation of amplitudes via Feynman path integrals and the construction of probabilities via Born rules represent methods of probing the topology of the given experimental setup. In fact the context independence of the prescriptions of quantum mechanics (i.e. Born rule and the rules for constructing amplitudes via summation over histories are independent of the specific experimental setup) represent a first expression of what I will call a "topologically co-variant" formulation of the laws of nature. If one decides to perform a quantum mechanical experiment, it appears that one is not allowed to assume a pre-existence of states, positions or paths between two measurable events. 
Moreover, one should not be allowed to assume the existence of one topological structure instead of another. The particular way in which space-time is connected is not observable a-priori. It becomes material only when an experiment that allows the inference of the topology is performed. So, at this moment we have general relativity, a theory that fails when dealing with "holes" in space-time (i.e. non-trivial topology) and quantum mechanics (i.e. a theory that is capable of probing a given topology). However, reasoning in the context of general relativity it appears plausible that to the topology of space-time should not be given an absolute character. As Einstein noted, only the "coincidences" [1] matter. In principle "connectedness" is whatever is in the context of quantum mechanics "the non-measurable in between-ness" so one is not supposed to know a-priori how this connectedness is realized. This is a common point in the reasoning of quantum mechanics and general relativity. It appears that in order to construct a quantum theory of gravity one has to pay attention to the way in which the laws of nature are expressed such that they do not depend on a particular topological choice. In my previous notes [6] I showed that several properties are to be considered relative if one makes use of several freedoms in defining the topology. It is my goal here to show that it is possible to construct a general theory of quantum gravity that is formulated explicitly independent of a particular topology. 
\section{ Independence of topology and the Universal Coefficient Theorem}
As argued in the previous chapter, the laws of physics should not depend on unobservable properties of spacetime. Specifically the choice of a particular coordinate system or a particular topology should not be relevant for the formulation of the laws of physics. I showed in a previous article [6] that specific choices of coefficient groups in cohomology may affect the observable connectedness of space-time (or generally of an abstract space) as measured by topological techniques. Here I focus on an aspect that maybe remained unspecified, namely what changes should be made in a theory in order for it to be independent on the way one choses to regard the topology? 
It appears to me that not all of the absolute character of the conventions related to spacetime has been removed while imposing diffeomorphism invariance. This appears to be what escaped Einstein in his formulation of general relativity and was unexpectedly introduced in the formulation of quantum mechanics, hence the apparent incompatibility of the two. 
In order to show what I mean let me start with the prescription of path integral quantization [4].  
R. Feynman observed that different functionals may give identical results when taken between any two states and argued that this equivalence between
functionals is the statement of operator equations in the language of path integrals. I assume that the standard prescription of computing quantum probabilities using 
quantum amplitudes is well known. If $P_{ac}$ is the quantum probability of measuring event $c$ when it follows the measurement of event $a$ then the probability must be
calculated as $P_{ac}=|\varphi_{ac}|^{2}$ where $\varphi_{ac}=\sum_{b}\varphi_{ab}\varphi_{bc}$ where the sum is over the possible intermediate states $b$ which, I emphasize, 
following Feynman (ref. [4], page 3 in manuscript) have no meaningful independent value. In a 1-space and 1-time dimensional context (the generalization to arbitrary dimensions should be straightforward) a succession of measurements may represent a succession of the space-coordinate $x$ at successive times $t_{1},t_{2},...$, 
where $t_{i+1}=t_{i}+\epsilon$. Let the observed value at $t_{i}$ be $x_{i}$. Classically the successive values
of $x_{1},x_{2},...$ define a path $x(t)$ when $\epsilon\rightarrow 0$. 
If the intermediate positions are actually measured one may talk about such a path with a well defined set of observed positions $x_{1},x_{2},...$ and the probability that 
the specified path $P(...x_{i},x_{i+1},...)$ lies in a region $R$ is given by the classical formula
\begin{equation}
 P=\int_{R}P(...x_{i},x_{i+1},...)...dx_{i}dx_{i+1}...
\end{equation}
where the integral is taken over the ranges of the variables which lie within the region $R$. 
If the intermediate positions are not measured then one cannot assign a value to them. In this case the probability of finding the outcome of a measurement in $R$ is
$|\varphi(R)|^{2}$ and $\varphi(R)$, i.e. the probability amplitude is calculated as
\begin{equation}
 \varphi(R)=\lim_{\epsilon\rightarrow 0}\int_{R}\Phi(...x_{i},x_{i+1},...)
\end{equation}
where $\Phi(...x_{i},x_{i+1},...)$ defines the path. In the given limit this object becomes a path functional. 
There should be no mystery nowadays that the probability amplitude should be calculated as 
\begin{equation}
 \varphi(R)=\lim_{\epsilon\rightarrow\ 0}\int_{R}exp[\frac{i}{\hbar}\sum_{i}S(x_{i+1},x_{i})]...\frac{dx_{i+1}}{A}\frac{dx_{i}}{A}...
\end{equation}
where $S$ is the action functional for the given path segment. 
In order to go a step further and define the wavefunction in this context I will continue to follow Feynman's paper [4]. The region $R$ considered above can be divided
into future and past with respect to a choice of a time position $t$. One can define the region $R'$ as the past and the region $R''$ as the future. 
The probability amplitude connecting these regions will be
\begin{equation}
 \varphi(R',R'')=\int \chi^{*}(x,t)\psi(x,t)dx
\end{equation}
where
\begin{equation}
 \psi(x_{k},t)=\lim_{\epsilon\rightarrow 0} \int_{R'}exp[\frac{i}{\hbar}\sum_{i=-\infty}^{k-1}S(x_{i+1},x_{i})]\frac{dx_{k-1}}{A}\frac{dx_{k-2}}{A}...
\end{equation}
and
\begin{equation}
 \chi^{*}(x_{k},t)=\lim_{\epsilon\rightarrow 0} \int_{R''}exp[\frac{i}{\hbar}\sum_{i=k}^{\infty}S(x_{i+1},x_{i})]\frac{1}{A}\frac{dx_{k+1}}{A}\frac{dx_{k+2}}{A}...
\end{equation}
In this way one can separate the ``past'' and the ``future'' via the functions $\psi$ and $\chi$.
One may also construct a closer equivalence to the matrix representation of quantum mechanics by introducing matrix elements of the form 
\begin{widetext}
\begin{equation}
<\chi_{t''}|F|\psi_{t'}>_{S}=\lim_{\epsilon\rightarrow\ 0}\int ... \int \chi^{*}(x'',t'')F(x_{0},...x_{j})exp[\frac{i}{\hbar}\sum_{i=0}^{j-1}S(x_{i+1},x_{i})]\psi(x',t')\frac{dx_{0}}{A}...\frac{dx_{j-1}}{A}dx_{j}
\end{equation}
\end{widetext}
In the limit $\epsilon\rightarrow 0$, $F$ is a functional of the path $x(t)$. 
At this moment one can define various equivalences between functionals. These are to be associated to operator equations in the matrix formulation. 
One can of course define $\frac{\partial F}{\partial x_{k}}$ and one can calculate the associated matrix element using an action functional $S$. 
Using the fact that the action functional appears as $exp(\frac{i}{\hbar} S)$ one obtains matrix equations as, say
\begin{equation}
 <\chi_{t''}|\frac{\partial F }{\partial x_{k}}|\psi_{t'}>_{S}=-\frac{i}{\hbar}<\chi_{t''}|F\frac{\partial S}{\partial x_{k}}|\psi_{t'}>_{S}
\end{equation}
which can be stated as a functional relation defined for an action $S$ as
\begin{equation}
 \frac{\partial F }{\partial x_{k}} \leftrightarrow -\frac{i}{\hbar}F\frac{\partial S}{\partial x_{k}}
\end{equation}
Using the fact that $S=\sum_{i=0}^{j-1}S(x_{i+1},x_{i})$ one can rewrite
\begin{equation}
  \frac{\partial F }{\partial x_{k}} \leftrightarrow -\frac{i}{\hbar}F[\frac{\partial S(x_{k+1},x_{k})}{\partial x_{k}}+\frac{\partial S(x_{k},x_{k-1})}{\partial x_{k}}]
\end{equation}
In the case of a simple 1-dimensional problem one can write 
\begin{equation}
\frac{\partial S(x_{k+1},x_{k})}{\partial x_{k}}=-m(x_{k+1}-x_{k})/\epsilon
\end{equation}
and
\begin{equation}
 \frac{\partial S(x_{k},x_{k-1})}{\partial x_{k}}=+m(x_{k}-x_{k-1})/\epsilon-\epsilon V'(x_{k})
\end{equation}
Neglecting terms of order $\epsilon$ one obtains
\begin{equation}
 m\frac{(x_{k+1}-x_{k})}{\epsilon}x_{k}-m\frac{(x_{k}-x_{k-1})}{\epsilon}x_{k} \leftrightarrow \frac{\hbar}{i}
\end{equation}
The important aspect here is that the order of terms in a matrix operator product corresponds to the order in ``time'' of the corresponding factors in a functional. 
The order of the factors in the functional is of no importance as long as the indexation of these factors is reflected in the ordering of the operators in the
matrix representation. This means the left-most term in the above equation must change order so that one obtains the well known commutation relation
\begin{equation}
 px-xp=\frac{\hbar}{i}
\end{equation}
\par These results should be nowadays generally known. One may observe that the indexation of the measurement outcomes, according to a time index (i.e. $\mathbb{Z}$-group), leads to the well known commutation relations. 
The ideas behind path integral quantization are kept intact when going to the relativistic context. However, when we have to go to a gravitational context the sum over geometries 
becomes non-trivial. In this sense one has to construct the (co)homology structure of the space and one has to deal with the universal coefficient theorem. This theorem states that 
a specific framework, constructed by the choice of a coefficient group in (co)homology is (up to (extension) torsion in (co)homology) equivalent with the choice of an integer coefficient 
group. However, some choices of coefficient groups may make some observables manifest while others may hide them. Moreover, simple order relations as the ones used in the proof
above are no longer uniquely defined. As I showed in my previous notes [6], what was identified by Feynman as a natural choice (time ordering) may in fact be just the result of a given coefficient group. 
In order to make the discussion more practical let me consider the formulation of Feynman's idea in the context of a space which is topologically non-trivial if looked upon via a coefficient group $G$. Let the probed space be $X$. In this case one cannot perform a trivial path-integral quantization due to the fact that the "hole in space" will make the path integral ill-defined. However, the universal coefficient theorem assures us that we can in principle chose another group structure for probing the space. Let this new structure be $G'$ such that one simply "overlooks" the "hole" in space under this group. I showed that this is possible in my previous notes [6].
In the context of Feynman's path integrals the paths will be indexed considering $G'=\mathbb{Z}_{n}$ case in which the succession of measurements will have the form $(x_{1},x_{2},...,x_{n},x_{1},...)$ but one will not exit a specific domain of the usual integers. One obtains a different topology when probing the space that contains a black hole. In fact, if one continues to use this group the black hole will essentially be invisible for an observer behind the horizon although he will never be able to exit it. The concepts must of course generalize for higher dimensional spaces but the idea is valid anyway. The construction of Feynman path integrals must be altered accordingly: 
\begin{equation}
 \varphi(R)=\lim_{\epsilon\rightarrow\ 0}\int_{R}[exp[\frac{i}{\hbar}\sum_{i}S(x_{i+1},x_{i})]...\frac{dx_{i+1}}{A}\frac{dx_{i}}{A}...]_{\mathbb{Z}_{n}}
\end{equation}
where the index shows that the sampling must be taken according to the new group structure. 
We face two distinct situations: one in which a massive black hole alters the topology of space-time and one in which all the diffeomorphism invariant properties are finite and there is no apparent problem. The topology-covariant formulations of the laws of physics are encoded in the statement of the universal coefficient theorem. Indeed the two situations are equivalent if one considers the $Ext$ and/or $Tor$ groups in (co)homology. In this sense homological algebra and the universal coefficient theorem are the mathematical tools required for the topological covariant formulation of the laws of physics somehow in the same way in which tensor calculus was the mathematical tool required for the general covariant formulation of the laws of physics. 
\par Now consider the exterior space also to be taken into account. In general there is no unique formulation of the universal coefficient theorem. One formulation that is particularly important is the following: given the tensor product of modules 
$H_{i}(X;\mathbb{Z})\otimes A$ one has the short exact sequence 
\begin{equation}
0\rightarrow H_{i}(X;\mathbb{Z})\otimes A \xrightarrow{\mu} H_{i}(X;A)\rightarrow Tor(H_{i-1}(X;\mathbb{Z}),A)\rightarrow 0
\end{equation}
Here $A$ is the alternative coefficient group, X is the analyzed space and $Tor$ is the torsion. 
More practically let for example $(C_{*},\partial)$ be a chain complex over a ring $R$. The chain groups are $C_{*}$. Then there is a map 
\begin{equation}
Hom_{R}(C_{q},M)\times C_{q} \rightarrow M
\end{equation}
that evaluates like
\begin{equation}
(f,z)\rightarrow f(z)
\end{equation}
This is a general formulation of a structure that has analogues in the covariant and contravariant structures in general relativity but also in the bra-ket notation of standard quantum mechanics. 
In quantum mechanics the amplitudes are characterized by complex numbers. The adjoint is defined naturally via hermitian conjugation giving rise to the bra-ket formalism and allowing the construction of theories preserving overall unitarity. 
In general relativity adjoints are constructed as dual 1-forms that appear as "covariant" indices and together with their contravariant counterparts assure that the theory can be formulated in a diffeomorphism invariant form despite the possible intrinsic curvature of spacetime. In principle the 1-forms take the value of a vector and produce a scalar. If $\tilde{P}$ is a 1-form and $\vec{V}$ is a vector then $<\tilde{P},\vec{V}>=\tilde{P}(\vec{V})=\vec{V}(\tilde{P})$. In the case of black holes there are well known and over-discussed issues related to the unitarity of the description also known under the generic name of "information paradoxes". These appear because the standard bra-ket construction does not map isomorphically into the (co)homology of the analyzed space and hence cannot be used in the same way. There are several ways in which possible pairings as the ones discussed above can be mapped into the realm of universal coefficient theorems. 
One possible pairing defined in the way described above is
\begin{equation}
< , >:H^{q}(C_{*};M)\times H_{q}(C_{*})\rightarrow M
\end{equation}
which relates homology with cohomology. 
This pairing is bilinear and its adjoint is a homomorphism 
\begin{equation}
H^{q}(C_{*},M)\rightarrow Hom(H_{q}(C_{*});M)
\end{equation}
Universal coefficient theorems, among other things, provide a measure of how this adjoint fails to be an isomorphism in terms of $Ext^{q}$ and $Tor_{q}$ [7]. 
Essentially what one has to observe is that in order to obtain the situation consisting of a black hole and an exterior asymptotically flat space-time one has to combine a structure with coefficients in $\mathbb{Z}_{n}$ with a structure with coefficients in $\mathbb{Z}$. The universal coefficient theorem puts restrictions on this combination. The map fails to be isomorphic and as a consequence not all observables that have a meaning under one choice of coefficients will have a well defined meaning in the other choice. Essentially it is the calculation of the $Ext$ and $Tor$ structures that tells us where the difference lies. It is only after one uses the universal coefficient theorem that one can talk meaningfully about observations made by an in-falling observer with respect to observations made by an observer standing at a large distance from the horizon. As the sequence is exact the observations of an in-falling observer and the observations of an observer standing far away will not share the same set of observables. Many "questions" asked by the far away observer will have no meaning for the in-falling one. This is the basic generalization of the invariance principles such that they include also the invariance with respect to a different choice of topology. The laws of nature must remain invariant under various choices of coefficient groups in (co)homology and this can be assured by the proper consideration of the $Ext$ and $Tor$ structures. 
There appears the question if there is some fundamental reason why I am using $\mathbb{Z}_{n}$ as a group structure instead of, say $\mathbb{Q}_{n}$ or $\mathbb{R}_{n}$. The simple answer is that nothing should stop me in using those other groups. As stated before there is no fundamental significance given to a choice of a group. One cannot associate an absolute topology to spacetime (be it discrete, continuous, etc.). Any such choice would be assimilable to the "ether" with respect to special relativity and would be immaterial. 
\section{conclusion}
The main discussion of this article revolves around the epistemological meaning of connectedness associated to a given space(time) and its fundamentally arbitrary nature. It is argued that in order to be able to formulate a theory that encompasses quantum mechanics and general relativity one must abandon the idea of absolute topological structure and rely on the universal coefficient theorem as on a "topology-covariant" formulation of the laws of physics.

\end{document}